# IEEE 802.11p-based Packet Broadcast in Radio Channels with Hidden Stations and Congestion Control[1]

Authors: Yunpeng Zang (*Member IEEE*), Bernhard Walke (*Fellow IEEE*), Guido Hiertz (*Member IEEE*), Christian Wietfeld (*Senior Member IEEE*)

Abstract: The Decentralized Congestion Control (DCC) algorithms specified in ETSI ITS standards [1] address the IEEE 802.11p MAC and provide reliability of periodic broadcast messages at high density of vehicles. However, the deterministic relation between controllable parameters, e.g. transmit power, frame duration, frame transmit rate and channel clear assessment threshold, and the effects of DCC algorithms, e.g. channel busy duration, frame interference-free reception probability and frame channel access delay, is still unknown since a correct mathematical analysis of the hidden station problem in CSMA networks is lacking. In this work, the hidden station problem in a linear IEEE 802.11p broadcast network is analyzed based on analytical results developed in [18] employing a modified MAC protocol model based on [3]. Simulation results validate the new analytical model for linear IEEE 802.11p networks w.r.t reliability and latency performances of Cooperative Awareness Message broadcast. Evidence is given that the model not only is valid for single-lane highways but also provides good approximate results for multi-lane highway scenarios. Our MAC layer analytical model of IEEE 802.11p broadcast reveals the quantitative relation between DCC parameters and congestion control effects in closed-form solution for linear vehicular networks.

## I. INTRODUCTION

Intelligent Transport Systems (ITS) improve safety and efficiency of road transport. Vehicular communication is an essential part of ITS and enables wireless information exchange among vehicles and to the road-side infrastructure. ITS applications such as Cooperative Awareness Service and Cooperative Autonomous Driving require reliable short range broadcast communication among vehicles. Protocol IEEE 802.11p provides Vehicle-to-Vehicle (V2V) and Vehicle-to-Infrastructure (V2I) communication for a range of up to 1 [km]. It specifies Medium Access Control (MAC) based on the Carrier Sense Multiple Access / Collision Avoidance (CSMA/CA) protocol but suffers from channel congestion at high vehicle densities

---

[1] This work has been submitted to the IEEE for possible publication. Copyright may be transferred without notice, after which this version may no longer be accessible.

[12]. Decentralized Congestion Control (DCC) algorithms have been developed in ETSI standards [1] to address this. DCC algorithms enforce each station to adjust its local transmission parameters like transmit power, packet duration, frame transmit rate and Channel Clear Assessment (CCA) threshold based on the channel load measured. However, the deterministic relation between these parameters and the performance of DCC algorithms, e.g. channel busy duration, frame interference-free reception ratio and frame channel access delay, is still unknown since a correct mathematical analysis of the hidden station problem in CSMA networks is lacking.

In [18] we introduce a methodology for modeling the hidden station problem in CSMA protocols and develop an analytical model to evaluate reliability and delay performance of one-dimensional (1-D) CSMA broadcast networks. The new hidden station model takes three input parameters: conditional channel access probability $p_{tx}$ at a station, frame duration $L$, and number of neighbors $R$ in single-side channel sensing range of a station, and gives closed-form solutions for MAC layer performance metrics including mean duration of channel busy period $T_{RB}$ whilst a station senses the channel continuously busy, interference-free probability $P_{IF}$ that a reception ends up with a frame free from any interference, and system goodput $G$ that is the fraction of overall time used for receiving interference-free frames at a station. Results from Monte-Carlo simulation show that the hidden station model provides accurate numerical results in linear CSMA networks [18].

In this paper we apply the hidden station model introduced in [18] for analysis of the IEEE 802.11p broadcast MAC protocol in a network with linear topology. To this end, we need to find the relation between the conditional channel access probability $p_{tx}$ at a station (introduced in [18]) and parameters of IEEE 802.11p, i.e. frame arrival rate $\lambda_F$ and minimum contention window size $CW_{min}$. This is achieved with an IEEE 802.11p protocol model originally introduced in [3], known as Bianchi's model, for analyzing throughout of the binary exponential back-off process of IEEE 802.11 Distributed Coordination Function (DCF). Bianchi's model is further modified in this paper for broadcast communication with non-saturated traffic load following the approach introduced in [11]. For a linear IEEE 802.11p network with hidden stations, our new model provides closed-form solutions for interference-free probability $p_{FIF}(d_{RX})$ of a frame reception as a function of topological distance $d_{RX}$ between the receiver and the transmitter, mean MAC layer channel access delay $\overline{D_S}$ of a frame, as well as system goodput $G$. Note, in a linear network the distance measured in units of stations is referred to as the *topological distance* in this study.

Our simulation results of Cooperative Awareness Message (CAM) [2] validate our analytical model in that it provides very good approximation to the simulated performance of CAM broadcast in realistic multi-lane highway scenarios, although the model is developed for a linear network topology only. The analytical model discovers the relation between performance degradation of CAM broadcast service and topological distance $d_{RX}$ between transmitter and receiver, and reveals the quantitative relation between both, reliability and delay performance, and controllable parameters like IEEE 802.11p contention window size $CW_{min}$, frame arrival rate $\lambda_F$, frame duration $L$, and the number of neighbors $R$ in single-side channel sensing range that is determined by the transmit power of stations. In this way, the new analytical model predicts the effects that DCC algorithms have on the performance of linear IEEE 802.11p networks with the presence of hidden stations.

This paper is outlined as follows: Section II reviews related work. Section III briefly reviews IEEE 802.11 DCF and IEEE 802.11p MAC protocol for broadcast communication. Section IV presents the system model. The analytical model for IEEE 802.11p broadcast communication is developed in Section V. In Section VI, we validate the new analytical model by results obtained from Monte-Carlo simulation. Section VII is dedicated to performance evaluation of ETSI CAM broadcast communication in a realistic multi-lane highway scenario using both, simulation and the new analytical model. Section VIII presents our conclusions.

## II.     RELATED WORK

Analytical models for IEEE 802.11 MAC protocol have been intensively studied. [3] provides an accurate analytical model for IEEE 802.11 DCF unicast communication and inspired many following work in this area. For example, [15] extends the model in [3] for supporting traffic with multiple priorities. [5] applies the model in [3] for analyzing the packet delay performance with non-saturated traffic load. [16] models broadcast communication with multiple priorities for safety-critical services in vehicular ad-hoc communications. Besides these efforts analysis of the hidden station problem in IEEE 802.11 networks is also covered, e.g. in [7], [10] and [6], to name some of them. [7] develops a frame work for modeling CSMA

MAC protocol in multi-hop networks. However, the solution in [7] is only given for specific network topology with limited number of hidden stations and traffic flows. [10] considers hidden stations in modeling IEEE 802.1p broadcast communication in a linear network based on [3]. The accuracy of the hidden station model in [10] is limited due to the size of the area containing potential hidden stations assumed to be constant, which is shown in [18] to be randomly distributed. [6] analyzes the broadcast performance of IEEE 802.11p in an infinite 1-D network. For simplicity reason, it is assumed that HSs transmit independently of each other according to a Poisson process. As shown in [17], this is a good approximation only if the conditional channel access probability $p_{TX}$ at each station is low.

A fundamental difference between our work and previous work on modeling the hidden station problem in IEEE 802.11 networks is that we decouple the hidden station problem from any specific MAC protocol and instead study the generic CSMA protocol in a given hidden station scenario [18]. This approach allows to directly apply the protocol model developed in [3] (representing the IEEE 802.11 binary exponential back-off process) to a hidden station scenario, so that effects resulting from hidden stations like prolonged channel busy time at a station, conditional channel idle probability, etc., can calculated. The conditional channel access probability $p_{tx}$ is the key parameter contained in both, the CSMA hidden station model and the IEEE 802.11 protocol model, as discussed in Section V.C.

### III. IEEE 802.11 DCF AND IEEE 802.11P BROADCAST

According to IEEE 802.11 DCF, when a station has a frame to transmit, the back-off entity first draws a random integer number $k$ for the back-off counter following uniform distribution in range $[0, CW_{min}]$. $CW_{min}$ is the minimum contention window size that usually takes the value $CW_{min} = 2^i - 1, i = 2, 3, 4 ...$ The back-off counter value decreases by one when the channel is sensed idle for a *Back-off Slot* $T_\sigma$. The back-off counter suspends when the channel is sensed busy and resumes only after the channel is sensed idle for a DCF Inter-Frame Space (DIFS) duration again. When its back-off counter reaches zero, a station starts transmission.

Unlike for unicast, IEEE 802.11 DCF for broadcast specifies neither acknowledgement (ACK) frame nor frame retransmission. Accordingly, the contention window size of a broadcast frame always stays at $CW_{min}$.

In case no frame is ready in the MAC queue to transmit after a frame transmission, a post-back-off is performed. In the post-back-off stage, the back-off entity draws a random number from $[0, CW_{min}]$ as the back-off counter value and starts to back-off, as described above for a frame that is ready for transmission. If a frame arrives in the MAC queue before the back-off counter reaches 0, the back-off entity goes on with the current back-off process and transmits the frame when the counter decreases to 0. In case the post-back-off counter reaches 0 with an empty queue, the back-off entity stops and waits for a new frame to arrive. After the post-back-off process, if a new frame arrives when the channel is sensed busy, the back-off entity draws a new back-off counter value and starts a normal back-off procedure. Otherwise, if the frame arrives and the channel is sensed idle, the frame is transmitted immediately without back-off.

IEEE 802.11p broadcast transmission applies the IEEE 802.11 Enhanced Distributed Channel Access (EDCA) rule. As far as broadcast transmission is concerned, EDCA is essentially DCF with extended support of up to four different channel access priorities, known as Access Categories (ACs). In this work we concentrate on the scenario with single AC with the presence of hidden stations. It is worth mentioning that the new analytical model can be extended to support multiple ACs with manageable efforts.

IEEE 802.11p stations that operate with 10 [MHz] channel spacing have physical layer (PHY) characteristics as given in Table 1

## IV. SYSTEM MODEL

Similar to the system model introduced in [18], the analytical model is developed for one-dimensional (1-D) network topology with an infinite number of stations, as shown in Figure 1. Stations have the same transmit power and receiver performance are assumed uniformly distributed with fixed spacing $1/\beta$ [m]. We use $R$ to denote the number of neighboring stations in the one-side channel sensing range $r$ of a station.

The time axis is assumed slotted with slot size $\sigma$ [s], which equals the length of IEEE 802.11p back-off slot $T_\sigma$. Stations are assumed time synchronized to the slot tact of the

Table 1: Characteristics of IEEE 802.11p with 10 [MHz] channel spacing

| Characteristics | Value |
|---|---|
| Backoff slot time ($T_\sigma$) | 13 [$\mu s$] |
| SIFS time ($aSIFSTime$) | 32 [$\mu s$] |
| DIFS time ($aDIFSTime$) | 58 [$\mu s$] |

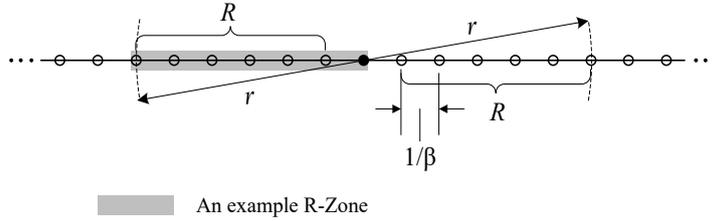

An example R-Zone

Figure 1:   Topology of infinite 1-D network with hidden stations

radio channel and to start transmission at the beginning of a slot. Frames are assumed to have constant length *L* counted in time slots.

Following [3], probability that the channel is sensed idle is assumed independent of the back-off procedure at a station. However, we have discovered (see Section VI.D) that this assumption becomes weak under non-saturated traffic load, if contention window size $CW_{min}$ of IEEE 802.11p is relatively small compared to frame length *L*.

Frame arrival at MAC layer of any station is assumed to follow a Poisson process with parameter $\lambda_F$ [frame/s].

For the IEEE 802.11p broadcast protocol model, in general, we assume each station has a MAC layer queue of infinite length, but for CAM broadcast we assume a maximum MAC layer queue length of 1, as explained in Section VII.A.

For easy adoption of the results of the hidden station model developed in [18], we assume each frame must perform at least one back-off counting-down event.

V.    ANALYTICAL MODEL FOR IEEE 802.11P BROADCAST

*A. Definition of IEEE 8021.11p Protocol Slots*

The concept of temporal spans in the protocol model for IEEE 802.11p is referred to as *protocol slot*. This definition of protocol slot uses the separation method for the *model Time slot* from [13] except that in this study at most one frame is transmitted after a back-off process. Figure 2 shows the protocol slots at station C, which is hidden to station A with respect to station B. Three types of protocol slots are defined, namely, idle protocol slot of length $T_{IP}$, busy protocol slot of length $T_{BP}$, and transmission (TX) protocol slot of length $T_{TP}$. All three types of protocol slots end with an idle IEEE 802.11p back-off slot, which is due to our assumption that each broadcast frame transmission is associated with a back-off procedure of non-zero counter value. In this model DIFS is treated as a part of the frame length *L* because after the channel returning to idle a station has to always wait for DIFS time before it starts counting down the backoff counter value. This results in $T_{TP} = L + \sigma$. As shown in Figure 2, the length of idle protocol slot is $T_{IP} = T_\sigma = \sigma$. Calculation of $T_{BP}$ is complicated due to hidden station problem that prolongs the busy protocol slot, e.g. the illustrated $T_{BP}$ at station B in Figure 2. In subsection V.C, we show how to calculate the mean value of $T_{BP}$ using results from the hidden station model developed in [18].

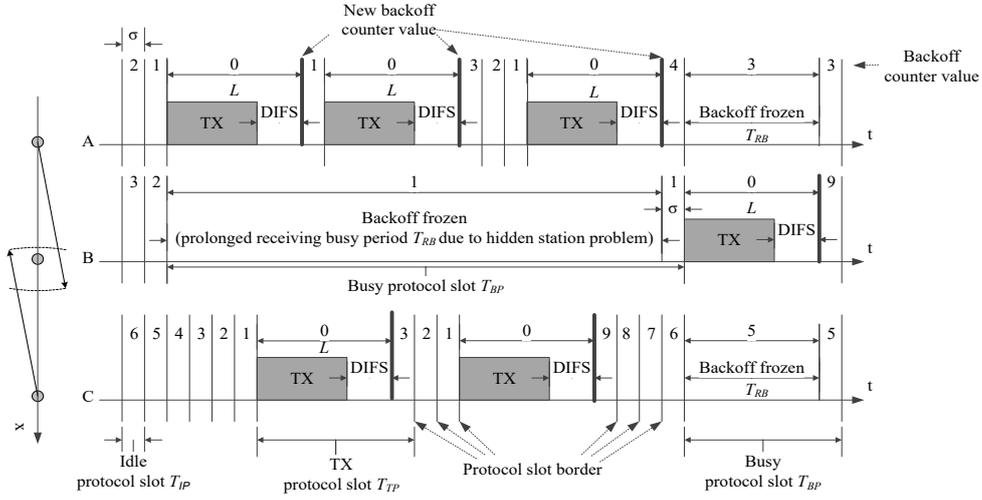

Figure 2: Protocol slots in the protocol model for IEEE 802.11p broadcast

B. *2-D Markov Chain for IEEE802.11p Broadcast Protocol*

With protocol IEEE 802.11p the probability that a station starts to transmit after sensing the channel as idle for a back-off slot $T_\sigma$ is determined by the back-off procedure described in Section III. A Markov chain model, shown in Figure 3, is introduced to model this back-off procedure. This Markov chain simplifies Bianchi's model [3] by keeping one back-off stage only for broadcast communication. This model also adopts the extension to the Bianchi's model, namely post-back-off stage for modeling non-saturated traffic load as proposed in [11], which is further improved in [8] with respect to the reset behavior of the post-back-off process.

All states and non-zero one step transitions of the 2-D random process $\{s(t), b(t)\}$ of the IEEE 802.11p protocol model are depicted in Figure 3. Transitions between states of the Markov chain occur at the border of protocol slots. States $\{0, k\}$, $0 \leq k \leq W - 2$, represent the status of a station with back-off counter value $k$ and at least one frame pending for transmission where $W = CW_{min} + 1$. States $\{0, k\}$, $0 \leq k \leq W - 2$, are collectively referred to as the *back-off stage*, whereas $\{-1, k\}$ are states in the *post-back-off stage*. There are $W - 1$ states in each stage because the case that an initial back-off counter value equals zero is assumed prohibited in this study, which prevents consecutive transmission of multiple frames after a single backoff process. Therefore, state {0,0} covers the situation that the initial back-off counter value equals one. Considering IEEE 802.11p broadcast is primarily designed for updating surrounding vehicles about the vehicle dynamics information of the transmitter, as we will discuss in VII.A, in practice it makes no sense for a vehicle to transmit consecutive messages after a single backoff process. Besides, our assumption has limited impact on modeling the channel access probability of IEEE 802.11p, particularly when the contention window size is relative large, e.g. $CW_{min} = 63$ or higher.

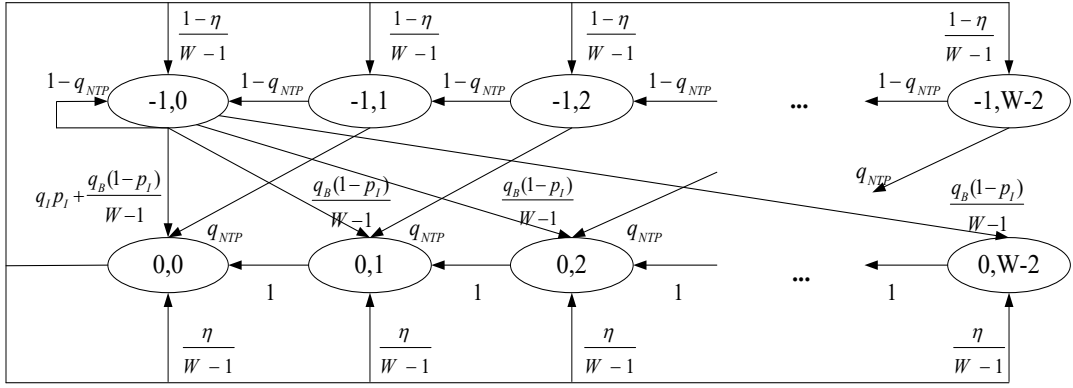

Figure 3: 2-D protocol Markov chain for IEEE 802.11p broadcast

$\eta$ is the probability that the MAC queue is not empty at the end of a TX protocol slot.

$q_{NTP}$ is the probability that at least one frame arrives during a non-TX protocol slot, i.e. during either an idle- or a busy- protocol slot.

$p_I$ and $1 - p_I$ are the conditional probabilities for the current protocol slot being idle and busy, respectively, conditioned on the station is not transmitting. Calculation of probability $p_I$ is complicated due to overlapped interferences from stations that are hidden to each other, e.g. the situation at station B in Figure 2. In Section V.C we show how to calculate $p_I$ using the hidden station model we have developed [18].

$q_I$ and $q_B$ are probabilities that at least one frame arrives during an idle and a busy protocol slot, respectively. Given $T_{IP}$ and $T_{BP}$ being the duration of idle and busy protocol slot, respectively, and frame arrival follows a Poisson process with arrival rate of $\lambda_F$, $q_I$ and $q_B$ are calculated as:

$$q_I = 1 - e^{-\lambda_F T_{IP}} \qquad (1)$$

$$q_B = 1 - e^{-\lambda_F T_{BP}} \qquad (2)$$

Given $p_I$, $q_I$ and $q_B$, the probability $q_{NTP}$ that at least one frame arrives during a non-TX protocol slot is

$$q_{NTP} = p_I \cdot q_I + (1 - p_I) \cdot q_B \tag{3}$$

It is worth mentioning that $q_I$, $q_B$, and $q_{NTP}$ calculated in (1) (2) (3), respectively, apply to any state in the IEEE 802.11p protocol model due to the assumed independence between channel idle probability $p_I$ and the back-off procedure, as originally introduced in [3].

The non-zero one step transition probabilities of the Markov chain are:

$$\begin{cases} P\{0,k|0,0\} = \dfrac{\eta}{W-1} & ,0 \leq k \leq W-2 \\ P\{-1,k|0,0\} = \dfrac{1-\eta}{W-1} & ,0 \leq k \leq W-2 \\ P\{0,k-1|0,k\} = 1 & ,1 \leq k \leq W-2 \\ P\{0,0|-1,0\} = q_I p_I + \dfrac{q_B (1-p_I)}{W-1} \\ P\{-1,0|-1,0\} = 1 - q_{NTP} \\ P\{0,k|-1,0\} = \dfrac{q_B (1-p_I)}{W-1} & ,1 \leq k \leq W-2 \\ P\{-1,k-1|-1,k\} = 1 - q_{NTP} & ,1 \leq k \leq W-2 \\ P\{0,k-1|-1,k\} = q_{NTP} & ,1 \leq k \leq W-2 \end{cases} \tag{4}$$

We adopt the short notation $P\{j_1, k_1|j_0, k_0\} = P\{s(t+1) = j_1, b(t+1) = k_1|s(t) = j_0, b(t) = k_0\}$ from [13].

Let $b_{j,k} = \lim_{t \to \infty} \Pr\{s(t) = j, b(t) = k\}$ be the limiting distribution of the 2-D Markov chain [13]. Owing to the chain regularities, the relation among all states can be written as:

$$\begin{cases} b_{-1,0} = b_{0,0} \cdot \dfrac{1-\eta}{W-1} \\ \qquad + b_{-1,1} \cdot (1 - q_{NTP}) + b_{-1,0}(1 - q_{NTP}) \\ b_{-1,k} = b_{0,0} \cdot \dfrac{1-\eta}{W-1} + b_{-1,k+1} \cdot (1 - q_{NTP}) & ,1 \leq k \leq W-3 \\ b_{-1,W-2} = b_{0,0} \cdot \dfrac{1-\eta}{W-1} \\ b_{0,0} = b_{0,0} \cdot \dfrac{\eta}{W-1} + b_{0,1} \\ \qquad + b_{-1,0} \cdot \left[ q_I \cdot p_I + \dfrac{q_B \cdot (1-p_I)}{W-1} \right] + b_{-1,1} \cdot q_{NTP} \\ b_{0,k} = b_{-1,0} \cdot \dfrac{q_B \cdot (1-p_I)}{W-1} \\ \qquad + b_{-1,k+1} \cdot q_{NTP} + b_{0,0} \cdot \dfrac{\eta}{W-1} + b_{0,k+1} & ,1 \leq k \leq W-3 \\ b_{0,W-2} = b_{-1,0} \cdot \dfrac{q_B \cdot (1-p_I)}{W-1} + b_{0,0} \cdot \dfrac{\eta}{W-1} \end{cases} \tag{5}$$

Following the convention in [3], $\tau$ is defined as probability that a station starts transmitting in a protocol slot. In this model $\tau$ equals the limiting probability $b_{0,0}$ of state $\{0,0\}$. From (5), we get the limiting distribution $b_{j,k}$ expressed using $\tau$ after imposing the normalizing condition:

$$1 = \sum_{j=-1}^{0} \sum_{k=0}^{W-2} b_{j,k} \quad (6)$$

From (5) and (6) we can express $\tau$ using $W$, $\eta$, $p_I$, $q_I$, $q_B$, and $q_{NTP}$:

$$\tau = \left[ (1-\eta) \cdot \frac{1}{q_{NTP}} + 1 + \frac{W-2}{2} \cdot [q_B \cdot (1-p_I)] \cdot \frac{1}{q_{NTP}} \right.$$
$$\cdot \frac{1-\eta}{W-1} \left[ 1 + \frac{1-q_{NTP}}{q_{NTP}} \cdot [1-(1-q_{NTP})^{W-2}] \right] + \frac{W-2}{2}$$
$$\cdot \eta + \frac{1-\eta}{W-1}$$
$$\cdot \left[ \frac{(W-3) \cdot (W-2)}{2} - \frac{1-q_{NTP}}{q_{NTP}} \cdot (W-3) + \left( \frac{1-q_{NTP}}{q_{NTP}} \right)^2 \right.$$
$$\left. \left. \cdot [1-(1-q_{NTP})^{W-3}] \right] \right]^{-1} \quad (7)$$

By imposing $\eta = 1$ to (7), i.e. the system is in saturated condition and all post-back-off states in the model are not reachable, $\tau$ is determined solely by parameter $W$. In this case, as already shown in [3], (7) reduces to

$$\tau = \frac{2}{W} \quad (8)$$

Note: Due to the assumption of no consecutive frame transmission after a single back-off process in this protocol model, we have $W$ in the denominator of (7), instead of $W+1$ as in [3].

Solving $\tau$ for unsaturated traffic load condition requires $p_I$, $q_{NTP}$ and $\eta$ to be solved first, which all depend on the length of busy protocol slot $T_{BP}$.

*C. Joint Solution for Hidden Station Model [18] and IEEE802.11p Protocol Model*

As shown in Figure 2, the prolonged busy protocol slot $T_{BP}$ at station B consists of a channel busy period $T_{RB}$ and a back-off slot $T_\sigma$. In [18] we show that the mean value of $T_{RB}$ can be solved using the hidden station model for CSMA protocols in a linear network with given parameters: $p_{tx}$ the channel access probability when a station senses the channel being idle for duration $\sigma$, $R$ the number of neighbor stations in the single-side channel sensing range, and $L$ the frame length (see Section D.1 of [18]). For the solution of $T_{RB}$ and $T_{BP}$, we need the relation between $p_{tx}$ and the limiting probability $\tau$ of state {0,0} in IEEE 802.11p protocol model. As illustrated in Figure 2, all three types of protocol slots end with an idle back-off slot time with duration $T_\sigma = \sigma$. The probability $p_{tx}$ that a station starts transmission after sensing the channel idle for duration $\sigma$ is exactly the probability that a station starts transmission in a protocol slot, i.e. the probability $\tau$ that the station is in state {0,0} of the Markov chain in Figure 2:

$$p_{tx} = \tau \qquad (9)$$

The probability $p_I$ that a station is in a idle protocol slot conditioned on that it is not in a TX protocol slot equals the probability that a station senses the channel idle again, conditioned on it does not start a transmission after having sensed the channel idle for a slot time $\sigma$. Using the time-domain Markov chain of the hidden station model, shown in Figure 6 of [18], we can calculate $p_I$:

$$p_I = \frac{p_{I,I}}{1 - p_{TX_{(L,1)},I}} \qquad (10)$$

Where, $p_{I,I}$ is the probability that the channel is idle in the next time slot $\sigma$ conditioned on the station has sensed the channel idle for a time slot $\sigma$, whereas $p_{TX_{(L,1)},I}$ is the probability that the station starts to transmit conditioned on the channel has been sensed idle for a slot time $\sigma$. Both $p_{I,I}$ and $p_{TX_{(L,1)},I}$ are given by the hidden station model in [18], more specifically in (6) of [18].

In order to find the solution to the IEEE 802.11p protocol model in hidden station scenarios, we treat each station as M/G/1 queuing system, where the back-off entity is modeled as the server, with the frame arrival rate $\lambda_F$ and service rate $1/D_S$. $D_S$ is the random variable representing the MAC layer service time of a frame. It is measured from the time when the frame reaches the head of the MAC queue, i.e. when it is marked as a ready frame for transmission, to the time when the TX protocol slot of this frame finishes. According to this definition, service time $D_S$ of an IEEE 802.11p broadcast frame consists of two parts, namely, the back-off time $T_{BK}$ and the TX protocol slot duration $T_{TP}$:

$$D_S = T_{BK} + T_{TP} \qquad (11)$$

According to the definition of TX protocol slot, $T_{TP}$ has a constant value:

$$T_{TP} = L + T_\sigma \qquad (12)$$

The value of $T_{BK}$ depends on two random variables: The number of back-off slots $K$, which is also the number of non-TX protocol slots experienced by a frame before its transmission, and the length of non-TX protocol slot $T_{NTP}$:

$$T_{BK} = K \cdot T_{NTP} \qquad (13)$$

In the following subsections, we calculate the mean value of $T_{BK}$ and solve the Markov chain in two different situations, namely saturated and non-saturated situations. Given $\rho$ is the system utilization factor of the queuing system, the system is said to be saturated if $\rho = 1$, and non-saturated if $\rho < 1$.

*C.1 Saturated System*

In a saturated system the MAC queue never goes empty and the value of $\eta$ in Figure 3 equals 1. In this case, the IEEE 802.11p protocol model consists only of back-off states $\{0, k\}$, $0 \leq k \leq W - 2$, leaving all post-back-off states $\{-1, k\}$, $0 \leq k \leq W - 2$, unreachable.

The number of non-TX protocol slots $K$ that a frame experiences before transmission is uniformly distributed in $[0, W - 2]$. Therefore, the mean back-off time $\overline{T_{BK}}$ is

$$\overline{T_{BK}} = \frac{W - 2}{2} \cdot \overline{T_{NTP}}, \tag{14}$$

where $\overline{T_{NTP}}$ is the mean duration of a non-TX protocol slot. The length $T_{NTP}$ of non-TX protocol slot takes values $T_\sigma$ with probability $p_I$ and $T_{BP}$, otherwise. Thus, $\overline{T_{NTP}}$ is

$$\overline{T_{NTP}} = p_I \cdot T_\sigma + (1 - p_I) \cdot \overline{T_{BP}}, \tag{15}$$

where $p_I$ is given in (10). The mean duration $\overline{T_{BP}}$ of a busy protocol slot is

$$\overline{T_{BP}} = \overline{T_{RB}} + T_\sigma \tag{16}$$

The value of $\overline{T_{RB}}$ in (16) and the values of $p_{I,I}$ and $p_{TX_{(L,1)},I}$ in (10) are solved by the hidden station model in [18], which requires knowledge of $p_{tx}$ besides the known parameters $L$ and $R$. To this end, (8) and (9) map the known parameter $CW_{min}$ to $p_{tx}$ in a saturated system:

$$p_{tx} = \frac{2}{W} = \frac{2}{CW_{min} + 1} \tag{17}$$

Provided $p_{tx}$ from (17) and known parameters $L$ and $R$, the hidden station model in [18], more specifically equation (20) in [18], gives the solution of $\overline{T_{RB}}$ and thus enables calculation of mean MAC layer service time $\overline{D_S}'$ of a frame in a saturated IEEE 802.11p system using equations (11) to (16) based on $L$, $R$ and $CW_{min}$.

Additionally, the minimal mean frame arrival rate $\lambda_F'$ that makes the system saturated can be calculated as

$$\lambda_F' = \frac{\rho}{\overline{D_S}'} = \frac{1}{\overline{D_S}'}, \tag{18}$$

where $\rho$ is the system utilization of the queuing system, whose value equals 1 in a saturated system, $\overline{D_S}'$ is the mean serving time of the queuing system, which equals the mean MAC layer service time of a frame in a saturated system.

*C.2 Non-saturated System*

In a non-saturated system, the distribution of the number $K$ of back-off slots, $0 \leq K \leq W - 2$, is more complicated due to the existence of post-back-off states, as shown in Figure 3. To solve for the mean service time $\overline{D_S}$ of a frame in a non-saturated system, we follow [4] and [5] and apply Probability Generating Functions (PGFs).

Based on (11) and (12) the PGF of $D_S$ is

$$D_S(z) = z^{L+1} \cdot T_{BK}(z) \tag{19}$$

where $T_{BK}(z)$ is the PGF of $T_{BK}$. Note, $T_\sigma = \sigma$ is the unit of time. According to (13) and the definition of PGF [9] we have

$$T_{BK}(z) = \sum_{k=0}^{W-2} Pr\{K = k\} \cdot [T_{NTP}(z)]^k, \tag{20}$$

where $T_{NTP}(z)$ is the PGF of $T_{NTP}$. According to (15) and (16), $T_{NTP}(z)$ is calculated as

$$T_{NTP}(z) = p_I \cdot z^{T_\sigma} + (1 - p_I) \cdot z^{T_B} = p_I \cdot z^{T_\sigma} + (1 - p_I) \cdot z^{T_\sigma} \cdot T_{RB}(z), \tag{21}$$

where $T_{RB}(z)$ is the PGF of the length of channel busy period $T_{RB}$ in the hidden station model in [18]. Inserting (20) and (21) into (19) we have

$$D_S(z) = z^{L+1} \cdot \sum_{k=0}^{W-2} Pr\{K = k\} \cdot [p_I \cdot z + (1 - p_I) \cdot z \cdot T_{RB}(z)]^k \tag{22}$$

The mean value (the first moment) of $D_S$ is calculated as the first derivative of $D_S(z)$ at $z = 1$

$$\begin{aligned}
\overline{D_S} &= D_S^{(1)}(z)|_{z=1} \\
&= (L + 1) \\
&\quad + \sum_{k=0}^{W-2} Pr\{K = k\} \cdot k \cdot \left[1 + (1 - p_I) \cdot T_{RB}^{(1)}(z)|_{z=1}\right] \\
&= (L + 1) + \sum_{k=0}^{W-2} Pr\{K = k\} \cdot k \cdot [1 + (1 - p_I) \cdot \overline{T_{RB}}],
\end{aligned} \tag{23}$$

where $\overline{T_{RB}}$ is given by the hidden station model (equation (20)) in [18]. $Pr\{K = k\}$, $0 \leq k \leq W - 2$, is the PMF of $K$ calculated as follows:

The value of $K$ depends on the state of back-off entity when the frame reaches the head of the MAC queue. As shown in Figure 3, state $\{0,0\}$ and states $\{-1, k\}, 0 \leq k \leq W - 2$, are the only states that the back-off entity can start to serve a new frame.

If a frame reaches the head of the MAC queue when the back-off entity is in state $\{0,0\}$, $K$ follows a uniform distribution in range $[0, W-2]$. The conditional distribution of $K$ in this case is:

$$Pr\{K = k \,|s_s = \{0,0\}\} = \frac{1}{W-1}, \qquad 0 \le k \le W-2 \tag{24}$$

$Pr\{s_s = \{0,0\}\}$ is the probability that the back-off entity is in state $\{0,0\}$, when it starts to serve the new frame. According to the protocol Markov chain in Figure 3, we have:

$$Pr\{s_s = \{0,0\}\} = \eta \tag{25}$$

If a frame reaches the head of the MAC queue when the back-off entity is in any of the states $\{-1, k+1\}, 0 \le k \le W-3$, $K$ takes the value of $k$ with probability 1:

$$Pr\{K = k | s_s = \{-1, k+1\}\} = 1 \quad, 0 \le k \le W-3 \tag{26}$$

From the protocol Markov chain, we have:

$$Pr\{s_s = \{-1, W-2\}\} = \frac{(1-\eta)}{W-1} \cdot q_{NTP} \tag{27}$$

And

$$Pr\{s_s = \{-1, k+1\}\} = \frac{(1-\eta)}{W-1} \cdot q_{NTP} \cdot \left[1 + \sum_{i=1}^{W-2-(k+1)} (1 - q_{NTP})^i \right], \tag{28}$$
$$0 \le k \le W-4$$

If a frame reaches the head of the MAC queue when the back-off entity is in state $\{-1,0\}$, $K$ has the conditional distribution

$$Pr\{K = k | s_s = \{-1,0\}\}$$
$$= \begin{cases} \dfrac{1}{q_{NTP}} \cdot \left[q_I \cdot p_I + \dfrac{q_B \cdot (1-p_I)}{W-1}\right] & ,k = 0 \\ \dfrac{1}{q_{NTP}} \cdot \dfrac{q_B \cdot (1-p_I)}{W-1} & ,1 \le k \le W-2 \end{cases} \tag{29}$$

$Pr\{s_s = \{-1,0\}\}$ is calculated as:

$$Pr\{s_s = \{-1,0\}\} = 1 - Pr\{s_s = \{0,0\}\} - \sum_{k=1}^{W-2} Pr\{s_s = \{-1, k\}\} \tag{30}$$

The PMF of $K$ is calculated by removing the condition in (24), (26) and (29), using (25), (27), (28), and (30), respectively, and combing the probabilities of the same $K$ value, for $0 \le K \le W-2$. (31) to (34) give the final results.

$$Pr\{K = 0\} = \frac{\eta}{W-1} + \frac{(1-\eta) \cdot q_{NTP}}{W-1}$$
$$\cdot \left[1 + \frac{(1-q_{NTP})(1-(1-q_{NTP})^{W-3})}{q_{NTP}}\right] + \frac{1}{q_{NTP}} \quad (31)$$
$$\cdot \left[q_I \cdot p_I + \frac{q_B \cdot (1-p_I)}{W-1}\right] \cdot Pr\{s_S = \{-1,0\}\}$$

$$Pr\{K = k\} = \frac{\eta}{W-1} + \frac{(1-\eta) \cdot q_{NTP}}{W-1}$$
$$\cdot \left[1 + \frac{(1-q_{NTP})(1-(1-q_{NTP})^{W-2-(k+1)})}{q_{NTP}}\right] + \frac{1}{q_{NTP}} \quad (32)$$
$$\cdot \left[\frac{q_B \cdot (1-p_I)}{W-1}\right] \cdot Pr\{s_S = \{-1,0\}\} \ , 1 \le k \le W-4$$

$$Pr\{K = W-3\}$$
$$= \frac{\eta}{W-1} + \frac{(1-\eta) \cdot q_{NTP}}{W-1} + \frac{1}{q_{NTP}} \cdot \left[\frac{q_B \cdot (1-p_I)}{W-1}\right] \cdot Pr\{s_S \quad (33)$$
$$= \{-1,0\}\}$$

$$Pr\{K = W-2\} = \frac{\eta}{W-1} + \frac{1}{q_{NTP}} \cdot \frac{q_B \cdot (1-p_I)}{W-1} \cdot Pr\{s_S = \{-1,0\}\}, \quad (34)$$

where $Pr\{s_S = \{-1,0\}\}$ is given in (30).

Using the frame arrival rate $\lambda_F$ and the mean service time $\overline{D_S}$ of the M/G/1 queuing system given in (23), we calculate the system utilization factor $\rho$ of the queuing system:

$$\rho = \lambda_F \cdot \overline{D_S} \quad (35)$$

On the other hand, according to its definition the system utilization $\rho$ of a queuing system is the mean fraction of time that the server is busy with ready work [9]. In IEEE 802.11p protocol model, we have

$$\rho = \frac{(b_0 - \tau) \cdot \overline{T_{NTP}} + \tau \cdot T_{TP}}{(1-\tau) \cdot \overline{T_{NTP}} + \tau \cdot T_{TP}}, \quad (36)$$

where $b_0$ is the limiting probability that the back-off entity is in any of states $\{0, k\}, 0 \le k \le$

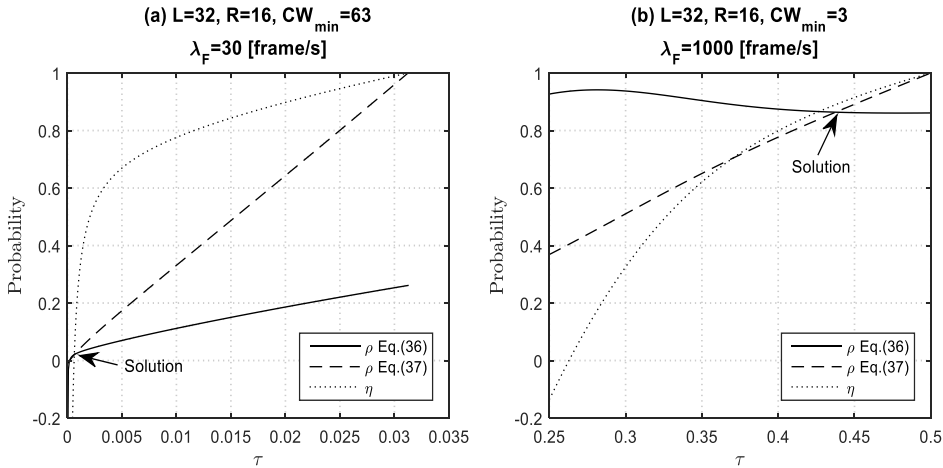

Figure 4: Numerical solution of the developed analytical model in non-saturated situation

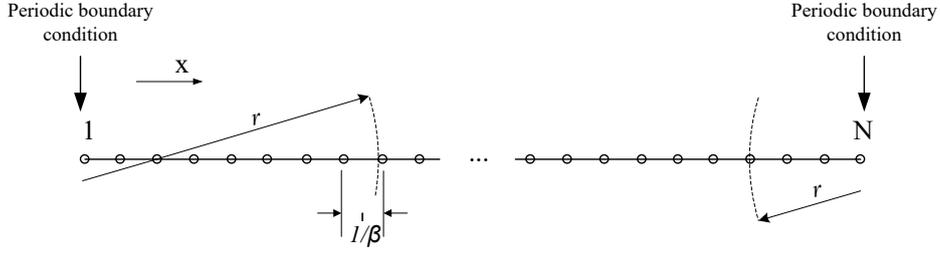

Figure 5:  Finite 1-D loop scenario for simulation

$W - 2$. Based on (5), $b_0$ is calculated as

$$b_0 = \sum_{i=0}^{W-2} b_{0,i} = 1 - \frac{\tau \cdot (1-\eta)}{q_{NTP}} \qquad (37)$$

The hidden station model and the protocol model in non-saturated system operation are solved by finding the value of $\tau$ in the protocol model, or equivalently the value of $p_{tx}$ in the hidden station model, which simultaneously satisfy (35) and (36):

$$\lambda_F \cdot \overline{D_S} = \frac{(b_0 - \tau) \cdot \overline{T_{NTP}} + \tau \cdot T_{TP}}{(1-\tau) \cdot \overline{T_{NTP}} + \tau \cdot T_{TP}} \qquad (38)$$

In this way, the performance of IEEE 802.11p broadcast, e.g. mean MAC service time of a frame, interference-free reception probability of a frame, and system goodput, in a linear hidden station scenario under non-saturated traffic load can be calculated with known parameters, $CW_{min}$, $\lambda_F$, $L$, and $R$.

Figure 4 shows two examples of the solution to the joint protocol and hidden station model, where the solution value of $\tau$ is found at the intersection of system utilization curves ($\rho$) calculated from (35) and (36), respectively. Another limiting condition is that $\eta$, the probability of non-empty queue at the end of each TX protocol slot, shall be in range $[0, 1]$ at the solution value of $\tau$. The parameter values used to calculate the results in Figure 4 are given in each sub-plot.

VI.  VALIDATION OF IEEE 802.11P ANALYTICAL MODEL

In this section, we aim to validate our analytical model using Monte-Carlo simulation results provided from a Matlab based simulator [17]. The simulation scenario is shown in Figure 5, where all stations are placed on a linear topology with constant spacing $1/\beta$. The channel sensing range $r$ of each station is far less than the length of the scenario. This simulation scenario avoids the border effect of any open-end topology by assuming periodical boundary condition. The simulation parameters of 1-D loop scenario are listed in Table 2:

## A. Essential Probabilities of IEEE 802.11p Protocol Model

Figure 6 shows probabilities $\tau$, $\eta$, $\rho$, and $p_I$ from (9), (7), (36) and (10), respectively, versus offered traffic load $\lambda_F$. Simulations are performed with identical setting of $L$ and $R$, whose values are given in the title of each subplot. The impact of contention window size is visible when comparing subplots with different $CW_{min}$ value. The analytical results for $\tau$, $\eta$, $\rho$, and $p_I$ closely match the simulated results in both subplots (a) and (b), under both saturated and non-saturated traffic load. This result validates the new analytical model. In subplot (a), i.e. with $CW_{min} = 3$, slight deviation between curves of the simulated and the analytical results are noticeable, when the offered traffic load $\lambda_F$ is between 100 [frame/s] and 1000 [frame/s]. This is a result of the small contention window size, as discussed in Section VI.D.

In both subplots of Figure 6, the values of $\tau$, $\eta$, and $\rho$ monotonically increase with increased traffic load $\lambda_F$ until system saturation ($\rho = 1$ at $\lambda_F$=1200 and 120 in subgraphs a and b, respectively). The performance of a saturated system does not depend on traffic load. Our analytical model calculates system performance under both, non-saturated and saturated operation. As shown in Figure 6, the system with a larger contention window size reaches saturation earlier. The probability $p_I$ that a station senses the channel idle in a non-TX protocol slot decreases with increased $\lambda_F$ value. If the value of $p_I$ reduces to zero, system goodput becomes null, as discussed in [18] about the system *synchronization point* and shown in Figure 8(a) of this paper. From Figure 6(b) one can see a larger $CW_{min}$ value can prevent the system from reaching the synchronization point and therefore always result in positive throughput due to the restricted value of $\tau$.

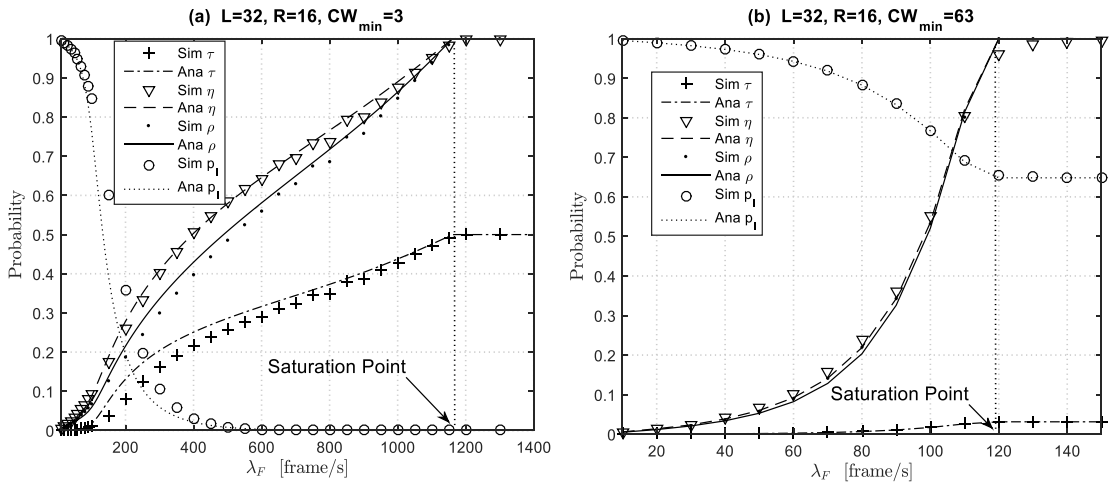

Figure 6:  Validated probabilities in the protocol model

## B. Time Metrics in IEEE 802.11p Protocol Model

Figure 7 shows time metrics of interest in the IEEE 802.11p protocol model, namely $\overline{T_{BP}}$, $\overline{T_{NTP}}$ and $\overline{D_S}$, whose analytical values are calculated from (16), (15) and (23), respectively. The close match of simulated and analytical results in subplot (b) with $CW_{min} = 63$ validates our model to calculate time metrics correctly. As mentioned in the previous subsection, remarkable deviation between the analytical results and the simulated results are visible for all time metrics in subplot Figure 7(a) for traffic load $\lambda_F$ between 100 [frame/s] and 1000 [frame/s]. We will discuss this effect in Section VI.D. Although analytical and simulated curves deviate from each other for small $CW_{min}$, both curves have a similar load dependent trend and the analysis may be accepted as a fair approximation.

Comparing subplots Figure 7(a) to Figure 7(b), one can see that a larger contention window size $CW_{min}$ results in longer mean MAC layer service time $\overline{D_S}$, whereas the mean duration of busy protocol slot $\overline{T_{BP}}$ and the mean duration of non-TX protocol slot $\overline{T_{NTP}}$ at a station are not affected by the value of $CW_{min}$. After saturation all time metrics stay at their values reached at the saturation point, which depends on the value of $CW_{min}$. It is worth noting that in a given hidden station scenario under non-saturated traffic load, time metrics such as $\overline{T_{BP}}$ and $\overline{T_{NTP}}$ from a station's point of view, are $CW_{min}$-independent, whereas the channel access delay $\overline{D_S}$, from a frame's point of view, depends on $CW_{min}$. The $CW_{min}$-independent performance of $\overline{T_{BP}}$ and $\overline{T_{NTP}}$ is a result of the statistically stationary channel access probability $p_{tx}$, which is protocol agnostic and is determined only by the traffic load and network topology, as far as the system is non-saturated. The $CW_{min}$-dependent performance of $\overline{D_S}$ is because of the mean back-off counter number $K$ in (13), whose value is determined by the contention window size $CW_{min}$, provided the value of $\overline{T_{NTP}}$ is independent of $CW_{min}$ in a non-saturated system.

Table 2: Simulation parameters of 1-D loop scenarios

| Parameters | Value |
|---|---|
| Number of stations ($N$) | 800 |
| Station density ($\beta$) | 1/30 [$station/m$] |
| Transmission probabilities ($\lambda_F$) | 10 ~ 1300 [$frame/s$] |
| Contention window Size ($CW_{min}$) | 3; 63 |
| Frame length ($L$) | 32 $\sigma$ |
| Number of one-side neighbors ($R$) | 16 |

## C. Interference-Free Reception Probability and System Goodput

The probability $p_{IF}$ of interference-free reception and the system goodput $G$ are calculated in Sections VI.D.2 and VI.D.3 of [18], respectively. Figure 8 shows the analytical results compared to simulation results. The accuracy of our analytical model is validated by a close match between analysis and simulation in subplots Figure 8(b). Still, the impact of the small value of $CW_{min}$ can be observed in subplot Figure 8(a), which will be discussed in Section VI.D.

Under non-saturated traffic load, with increased traffic load the system goodput $G$ reaches a maximum at $\lambda_F = 60$ [frames/s] and decreases down to a constant value determined at the saturation point, whereas the interference-free probability $p_{IF}$ monotonically decreases with increased traffic load till the saturation point is reached. Similar to time metrics $\overline{T_{BP}}$ and $\overline{T_{NTP}}$ discussed in the previous section, $p_{IF}$ and $G$ do not depend on $CW_{min}$ in a non-saturated system.

## D. Impact of Small Contention Window Size on Stationarity of Conditional Channel Access Probability of IEEE 802.11p

As discovered in the previous subsections, deviation between analytical and simulation results are noticeable for all performance metrics if the contention window size $CW_{min}$ is small, e.g. $CW_{min} = 3$ as shown in Figure 7(a). Then, analytical results provide a lower bound to system performance, as shown in Figure 8(a) for $p_{FIF}$ and $G$, respectively.

As discussed in Section 6.7.4 of [17], this effect results from the assumption of stationary and back-off state independent conditional channel idle probability $p_I$ (originally introduced in

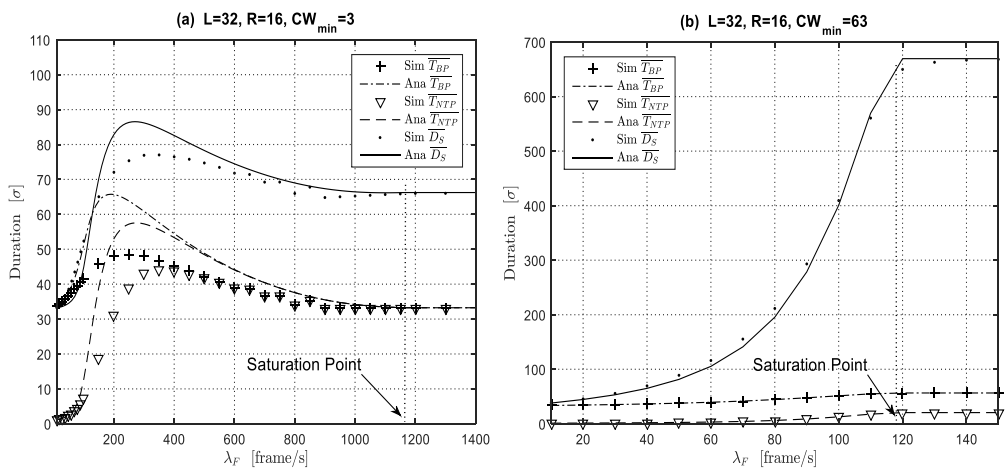

Figure 7: Time metrics in IEEE 802.11p protocol model

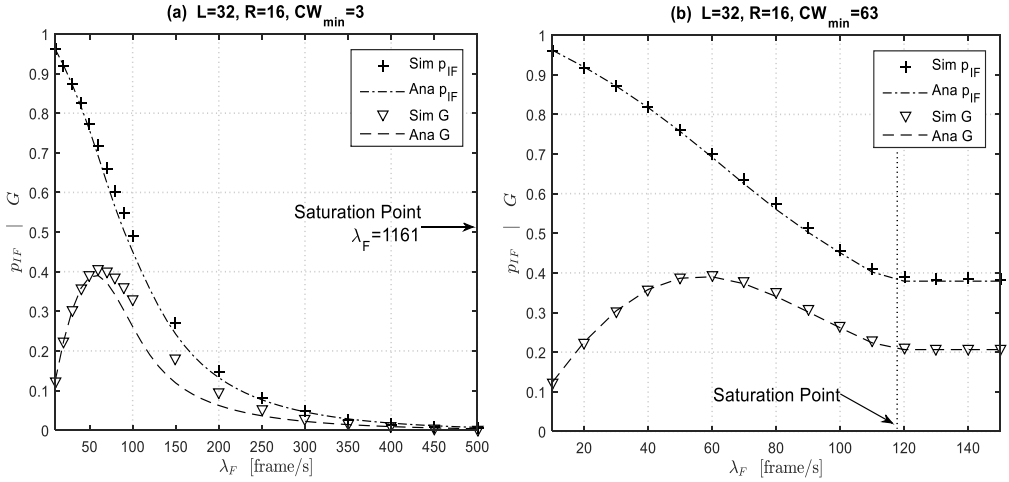

Figure 8: Probability of interference-free reception burst and system goodput

[3]) that does not hold for a small $CW_{min}$ value (relative to the frame size $L$, both counted in $\sigma$) in a non-saturated system. [17] also discovers that conditional channel idle probability $p_I$ depends on back-off state in a non-saturated system, where higher values of $p_I$ are observed in post-back-off states $\{-1, k\}$, $0 \leq k \leq W - 2$, in comparison with back-off states $\{0, k\}$, $0 \leq k \leq W - 2$, other than assumed in our model that follows [3] in this aspect.

## VII. PERFORMANCE ANALYSIS OF CAM BROADCAST IN IEEE 802.11P NETWORK

Cooperative Awareness Messages (CAMs) are defined in standard ETSI EN 302 637-2 [2] to disseminate by vehicles timely status and attribute information, e.g. position, speed, and heading, from a vehicle. CAMs serve as the basis of many vehicular safety services, e.g. the Cooperative Awareness Service (CAS), through which each vehicle tracks the status of the surrounding vehicles. CAMs are periodically broadcasted at each vehicle with a frequency between 1 and 10 [Hz] [2].

The MAC layer performance of CAS can be evaluated using the CAM update interval $T_{UI}$ at a receiver. $T_{UI}$ is defined as the time between two consecutive receptions of interference-free CAM frames from the same transmitter. Particularly, we define $T_{UI}(d_{RX})$ to be the update interval of CAM at a receiver having a topological distance $d_{RX}$ to the transmitter. Another performance metric of CAM, from a frame's point of view, is the frame interference-free probability $p_{FIF}(d_{RX})$. $p_{FIF}(d_{RX})$ is defined as the probability that a CAM frame is received without being interfered, given the topological distance between transmitter and receiver is $d_{RX}$.

In this section, we apply the IEEE 802.11p broadcast model developed in Section V to analyze the performance of CAM broadcast. Both $T_{UI}(d_{RX})$ and $p_{FIF}(d_{RX})$, with $1 \leq d_{RX} \leq R$, are studied in a linear network topology with respect to the density of stations, the size of the CAM frame, and the stations' information update frequency. Further, for reaching optimal $T_{UI}(d_{RX})$ and $p_{FIF}(d_{RX})$ performance congestion control mechanisms, e.g. transmit power control, transmit data rate control and frame duration control are also discussed for realistic multi-lane highway scenarios.

*A. Modeling CAM Broadcast in IEEE 802.11p Network*

As a CAM carries the up-to-date information of the transmitting station, there is no need to keep any old frame at the MAC layer queue, if a new CAM frame with updated information arrives. Therefore, a new CAM frame always overwrites the old one in the queue [6]. Instead of an infinite MAC queue length, we introduce a queue length of 1. This does neither change the way CSMA stations interact with each other in a hidden station scenario, nor affect the back-off procedure of the IEEE 802.11p. Thus, the IEEE 802.11p broadcast model developed in the previous section is applicable to CAM broadcast analysis. The effect of MAC layer queue length 1 is modeled by $\eta$, the probability that at least one frame is in the queue at the end of a TX protocol slot:

$$\eta = 1 - e^{-\lambda_F \cdot T_\sigma}, \tag{39}$$

where the mean frame arrival rate $\lambda_F$ equals the information update frequency of CAS, and $T_\sigma$ is the duration of a back-off time slot. (39) is based on the observation that with queue length 1 the value of $\eta$ equals the probability that at least one frame arrives during the back-off time slot $T_\sigma$ that follows the frame transmission in a TX protocol slot.

According to (39) the value of $\eta$ reaches 1 only if $\lambda_F = \infty$, i.e. the system reaches saturation only at an infinite frame arrival rate, since any frame arriving at a non-empty queue replaces the queued frame and does not contribute to the traffic load of the system. Accordingly, (35) for calculating the system utilization $\rho$ is no longer valid. Instead, we solve for the value of $\tau$ using the equation system of (7) and (39) containing known parameters $L$, $R$, $CW_{min}$, and $\lambda_F$ and unknowns $q_{NTP}$, $q_B$, $q_I$, and $p_I$ that are functions of $\tau$, as discussed in section V.

## B. Mean CAM Update Interval within Channel Sensing Range r

By definition the value of CAM update interval $T_{UI}(d_{RX})$, $1 \leq d_{RX} \leq R$, is measured as the time between two consecutive interference-free receptions from the same transmitter. Therefore, the mean value of $T_{UI}(d_{RX})$ at a receiver is calculated using the mean duration $\overline{T_{RXP}}$ between the starting time of two consecutive reception events, the mean interference-free probability of a reception $p_{IF}$, and the conditional distribution $f_{d_{RX}|IF}$ of $d_{RX}$ given a reception is free from interference:

$$\overline{T_{UI}(k)} = 2 \cdot \left[\frac{1}{\overline{T_{RXP}}} \cdot p_{IF} \cdot f_{d_{RX}|IF}(k)\right]^{-1}, 1 \leq k \leq R \quad , \tag{40}$$

where $\overline{T_{RXP}}$, $p_{IF}$, and $f_{d_{RX}|IF}$ are given by the hidden station model in (5.35), (5.48), and (5.49) of [17], respectively. Factor 2 in the right part of (40) is because the calculation of $\overline{T_{RXP}}$ in (5.35) of [17] only considers frames received from one side of the station. According to the assumption of homogeneous behavior of all stations in Section IV, the value of $\overline{T_{UI}(d_{RX})}$ for stations at the left- and right-side of the receiver shall be equal for a certain $d_{RX}$ value in $1 \leq d_{RX} \leq R$.

## C. Interference-Free Probability of Received Frame

The mean CAM update interval $\overline{T_{UI}(d_{RX})}$, $1 \leq d_{RX} \leq R$, can be also calculated from a frame's point of view using mean transmission period $\overline{T_{TXP}}$ at the sender (referred to as the *concerned station* in the following discussion), the conditional probability $p_{ASYNC}(d_{RX})$, $1 \leq d_{RX} \leq R$, that the station at topological distance $d_{RX}$ does not start to transmit at the same time as the concerned station does, and the conditional frame interference-free probability $p_{FIF}(d_{RX})$, at the receiver at topological distance $d_{RX}$:

$$\overline{T_{UI}(d_{RX})} = \left[\frac{1}{\overline{T_{TXP}}} \cdot p_{ASYNC}(d_{RX}) \cdot p_{FIF}(d_{RX})\right]^{-1}, 1 \leq d_{RX} \leq R \tag{41}$$

where $\overline{T_{TXP}}$ is given in (18) of [18]. As explained in the hidden station model in [18], in a linear CSMA network the number $d_F$ of consecutive stations that simultaneously sense channel idle follows a geometric distribution with parameter $p_{O,F}$, (see Section VI.B1 of [18]). Thus the probability that a station at topological distance $d_{RX}$ from the concerned station starts to transmit in the next time slot conditioned on the concerned station also starts transmit in the next time slot is $(1 - p_{O,F})^{d_{RX}} \cdot p_{tx}$, where $p_{tx}$ is the channel access probability of a station that equals $\tau$ in the IEEE 802.11p protocol model, as discussed in Section V.C. Therefore,

$$p_{ASYNC}(d_{RX}) = 1 - (1 - p_{O,F})^{d_{RX}} \cdot p_{tx} \tag{42}$$

In Section VII.A we outline the CAM broadcast model for solving $\tau$ or $p_{tx}$ with given parameters $L$, $R$, $\lambda_F$ and $CW_{min}$, and $p_{O,F}$ is solved using the hidden station model with known parameters $L$, $R$, and $p_{tx}$, as discussed in Section V.C of [18].

Given $\overline{T_{UI}(d_{RX})}$, $1 \leq d_{RX} \leq R$, is calculated with (40), we can get the frame interference-free probability $p_{FIF}(d_{RX})$ by rewriting (41) as follows:

$$p_{FIF}(d_{RX}) = \frac{\overline{T_{TXP}}}{\overline{T_{UI}(d_{rx})} \cdot p_{ASYNC}(d_{RX})} \quad , 1 \leq d_{RX} \leq R \quad (43)$$

Note, factor 2 in (40) does not exist in (41) and (43), because $d_{RX}$ can only be on either left- or right-side of the concerned station in the calculation of $\overline{T_{TXP}}$, $p_{ASYNC}(d_{RX})$, and $\overline{T_{UI}(d_{RX})}$, the last of which is clarified by (40).

*D. Performance Analysis of CAM in Multi-lane Highway Scenarios*

The CAM broadcast model is also checked against simulation results in terms of CAM update interval $T_{UI}(d_{RX})$, $1 \leq d_{RX} \leq R$, and frame interference-free probability $p_{FIF}(d_{RX})$, $1 \leq d_{RX} \leq R$.

The simulation scenario is a six-lane straight highway with periodic boundary conditions, as shown in Figure 9. This highway has three lanes in each direction and a median strip separating opposite traffic lanes. As marked in Figure 9, the width of a lane and the median strip is 5 [m] and 2 [m], respectively. Vehicles' location in the scenario is a static snapshot of the simulated free-flow traffic using the microscopic car-following Intelligent Driver Model (IDM) [14]. The details about IDM and the traffic simulation are given in Appendix E of [17]. Table 3 summarizes key parameters of the simulated multi-lane vehicular network.

In multi-lane highway scenarios, there exist two or more vehicles having identical neighbor vehicles in their channel sensing range, e.g. vehicle *A* and *B* in Figure 9, which are located closely in the x-coordinate but on different lanes. Vehicle *A* and *B* suffer from the same set of hidden stations and are modeled as one station in the analytical model but are treated separately in the simulation model. For this same reason, the number of vehicles $R'$ in one side channel sensing range of a vehicle is not exactly the number of hidden stations $R$ in the analytical model. However, results shown in Figure 10 and Figure 11 indicate the effect from this inconsistency is minor and the accuracy of the analytical results derived using $\lceil \bar{R} \rceil$ is acceptable for the given parameter values in Table 3, where $\lceil \bar{R} \rceil$ takes the first integer that is greater or equal to the mean value of $R$ at all stations.

Table 3: Parameters of multi-lane vehicular network

| Characteristics | Value |
|---|---|
| Number of vehicles | 800 |
| Vehicle density ($\beta$) | 0.11 [vehicle/m] |
| Channel sensing range ($r$) | 184.60 [m] |
| Mean number of vehicles in one side channel sensing range ($\overline{R'}$) | 20.75 |
| Mean number of stations in one side channel sensing range ($\overline{R}$) | 15.98 |

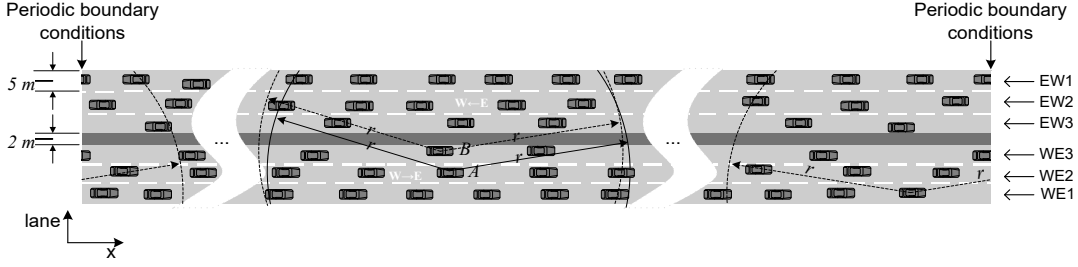

Figure 9: Multi-lane highway scenario for validating CAM model

Protocol settings are identical to Section VI, except that the length of the MAC layer queue is set to 1 for simulating the CAM broadcast.

As the major goal is to validate the new MAC layer model for CAM broadcast in vehicle network on multi-lane highway, we do not consider the impacts of fading channel and physical layer receiver sensitivity. Instead, all performance metrics are based on the interference-free probability, which is considered more stringent than physical layer frame success probability.

*D.1 Mean CAM update interval*

Calculation of $\overline{T_{UI}(d_{RX})}, 1 \leq d_{RX} \leq R$, using (40) is compared to simulation results, as shown in Figure 10 where scenario settings are given in the title of each sub-plot. For investigating mean CAM update interval as a function of $d_{RX}$ we study three values of $\lambda_F$, namely 10, 40, and 60 [frame/s].

In both subplots a) and b) analytical and simulated results are close to each other, except for a relatively large deviation in subplot b) for $\lambda_F = 60$ [frame/s] at $d_{RX} \geq 8$, which is a consequence of $R' \neq R$.

In both subplots, with a low frame arrival rate $\lambda_F$, e.g. $\lambda_F = 10$ [frame/s], $\overline{T_{UI}(d_{RX})}$ increases much slower along with increased value of $d_{RX}$, compared to the curves for higher values of $\lambda_F$, e.g. $\lambda_F = 40$ [frame/s] and $\lambda_F = 60$ [frame/s]. The value of $\overline{T_{UI}(1)}$ at $d_{RX} = 1$ with $\lambda_F = 10$ [frame/s] is much larger than that with $\lambda_F = 40$ [frame/s] and $\lambda_F = 60$ [frame/s]. The large $\overline{T_{UI}(1)}$ value at a low $\lambda_F$ value is attributed to less transmission events under a low value of $\lambda_F$. The steep curve of $\overline{T_{UI}(d_{RX})}$ at a high $\lambda_F$ value suggests that in a hidden station scenario, a receiver having closer distance to the transmitter has higher probability of receiving an interference-free frame compared to a receiver located farther away. This trend becomes more obvious when traffic load at each station increases. It is worth mentioning that in a homogenous system, as studied here, this effect unbiasedly applies to every station as a CAM transmitter and all its neighbors as receivers.

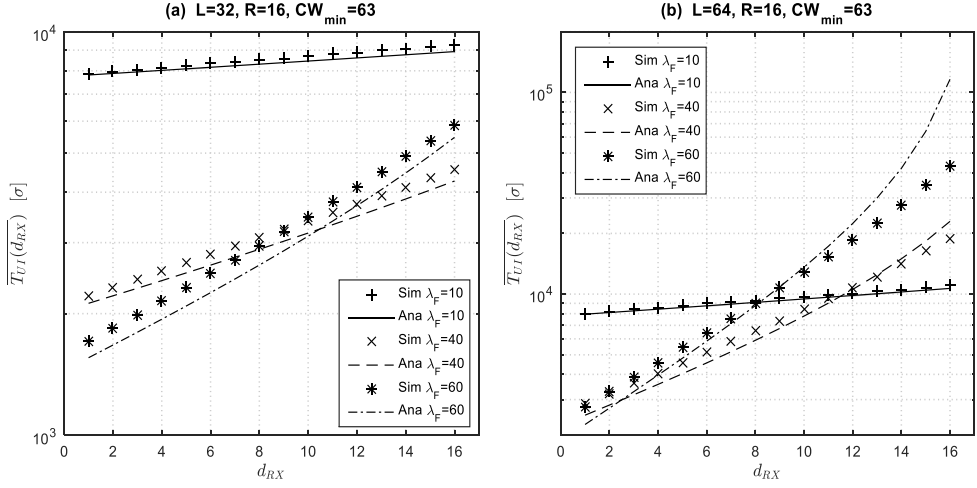

Figure 10: Mean CAM update interval $\overline{T_{UI}(d_{RX})}$

Time metrics in the CAM broadcast are sensitive to frame size. This explains the generally worse $\overline{T_{UI}(d_{RX})}$ performance with a larger frame size $L = 64$ in Figure 10(b), compared to Figure 10(a), where $L = 32$.

### D.2 Interference-Free Probability $p_{FIF}$ of Received Frame

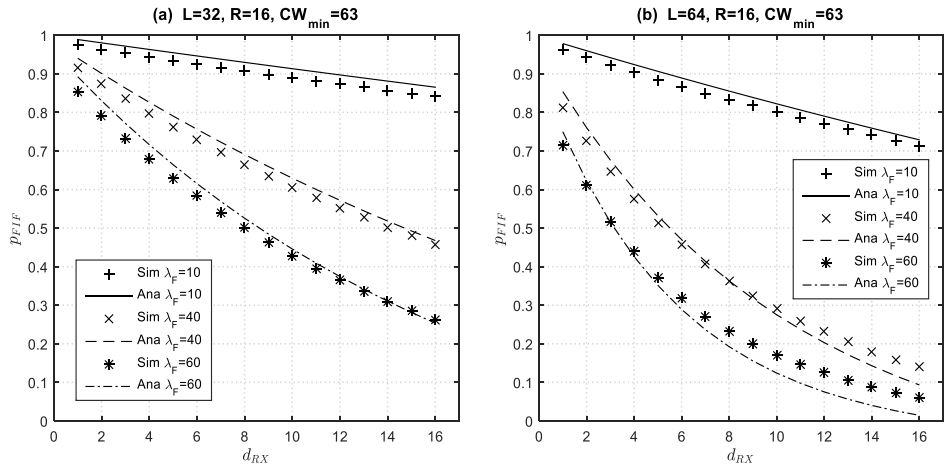

Figure 11: Interference-free probability $p_{FIF}$ of received frame at $d_{RX}$

Figure 11 compares calculated results for $p_{FIF}(d_{RX}), 1 \le d_{RX} \le R$ from (43) and simulated results. Three $\lambda_F$ values are studied to check applicability of our analytical model for calculation of $p_{FIF}(d_{RX})$ to real-world problems. As shown in Figure 11, simulated results closely match analytical results. By comparing Figure 11(a) to Figure 11(b) one can see the negative impact of large frame size L on $p_{FIF}(d_{RX})$ performance. The results also confirm our expectation that a frame has a higher probability of being received interference-free at a closer by receiver than at a farther distant receiver. Besides, at all $d_{RX}$ values a higher traffic load at each station always results in lower $p_{FIF}(d_{RX})$.

*D.3 Performance optimization of CAM broadcast*

In this section, the MAC layer performance of CAM broadcast in IEEE 802.11p vehicular networks is evaluated using the new CAM broadcast model.

Having the cooperative awareness service (CAS) in mind, we choose the value of CAM update interval $\overline{T_{UI}(d_{RX})}$ at $d_{RX} = 8$ and the maximum possible value of $d_{RX}$ in each scenario as the performance metric. This is because, on the one hand, the results in Section VII.D.1 show a monotonically decreasing performance of $\overline{T_{UI}(d_{RX})}$ with increased $d_{RX}$, on the other hand, for safety services information from a vehicle that is in close vicinity of a receiver, i.e. with a low value of $d_{RX}$, is far more important than that from a vehicle that is located at farther distance, i.e. with a high value of $d_{RX}$. In a highway scenario, the value $d_{RX} = 8$ in the driving direction and the reverse direction of a receiving vehicle shall cover the directly adjacent vehicles driving in front of and behind the concerned vehicle in the same lane, as well as in the neighbor lanes.

In order to be comparable to the specification of the CAM standard [2], analytical results of the CAM update interval performance shown in Figure 12 and Figure 13 are presented using the units derived according to the parameters in Table 1 and Table 4. Besides, we change the unit of $\lambda_F$ from [frame/s] to [Hz] to reflect that a new CAM message generated at the transmitter is essentially an information update of the vehicle.

Table 4:  Calculation parameters for CAM broadcast

| Parameter Setting | Value |
| --- | --- |
| IEEE 802.11p PHY mode | QPSK1/2 |
| CAM frame size $P_S$ | 200 [B]; 512 [B] [*] |
| Vehicle density $\beta$ | 0.2 [vehicle/m] |
| Channel sensing range $r$ | 20 [m] ~ 640 [m] |

[*] With PHY mode QPSK1/2, $P_S = 200$ [B] corresponds to $L = 32$; $P_S = 512$ [B] corresponds to $L = 64$.

Figure 12 shows the analytical results of $\overline{T_{UI}(d_{RX})}$ for frame size $P_S = 200$ [B] and $P_S = 512$ [B] with $CW_{min} = 63$ and different channel sensing range, as indicated in the title of each sub-plot. Figure 12(a) and Figure 12(b) are for the results of each frame size, respectively.

Due to the increasing number of vehicles in the channel sensing range $r$ of each vehicle, a deteriorating performance of $\overline{T_{UI}(d_{RX})}$ is observed with increased value of $r$, for both frame sizes. Nevertheless, given that each vehicle updates its information with a frequency not lower than 1 [Hz], the performance of $\overline{T_{UI}(8)}$ in all investigated scenarios is below 1 s, as shown in Figure 12. Here, the largest channel sensing range $r = 640$ [m] corresponds to 128 sensible neighbors at each side, either ahead or behind, of a vehicle. As expected, the performance at the maximum $d_{RX}$ in each scenario is far worse than that at $d_{RX} = 8$. Particularly, with $P_S = 512$ [B] and $r = 640$ [m] the value of $T_{UI}(128)$ is hardly below 1 s, regardless of the offered traffic load.

As shown in Figure 12, by increasing the value of $\lambda_F$ beyond 1 [Hz], we get better performance of $\overline{T_{UI}(d_{RX})}$ towards the optimal value in all scenarios. After the optimal value, which is at different value of $\lambda_F$ for different value of $r$, the performance of $\overline{T_{UI}(d_{RX})}$ deteriorates in all scenarios with a gradually decreasing rate, when the value of $\lambda_F$ further increases.

The results for $\overline{T_{UI}(d_{RX})}$ with a larger frame size, e.g. $P_S = 512$ [B] in sub-plot Figure 12(b), are generally worse compared to that in Figure 12(a), which has a smaller frame size $P_S = 200$ [B]. This confirms the discussion in Section VII.D.1 about the negative impact of the frame size on the time metrics in CAM broadcast.

Figure 13 gives the analytical results with the same settings used for Figure 12, except for a larger contention window size of the IEEE 802.11p system, i.e. $CW_{min} = 127$. The difference between Figure 12(a) and Figure 13(a), as well as the difference between Figure 12(b) and Figure 13(b), respectively, clearly show the effectiveness of a larger $CW_{min}$ value in limiting channel access probability at each station and, thus, in preventing further degradation of the $\overline{T_{UI}(d_{RX})}$ performance with increased value of $\lambda_F$

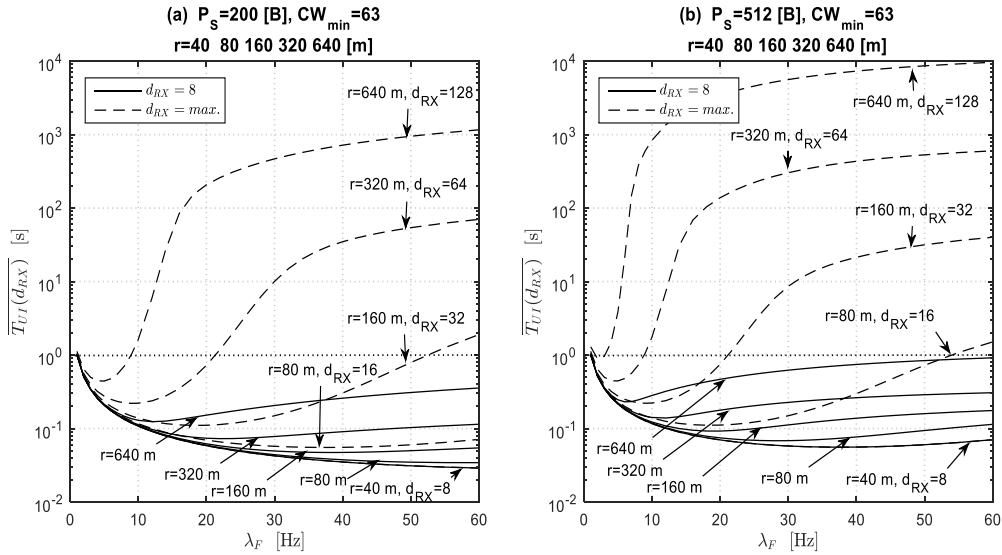

Figure 12: CAM update interval evaluated at $d_{RX}$=8 and $d_{RX}$=max. with $CW_{min}$=63

The analytical results for $\overline{T_{UI}(d_{RX})}$ provide the following insights into the control mechanisms towards optimal performance of CAM broadcast in hidden station scenarios:

Keeping the number of stations $R$ in the channel sensing range $r$ of each station below a certain value, e.g. $R = 128$ for the investigated scenarios, is the most effective way of avoiding the hidden station problem. For a given receiver sensitivity level, this can be implemented using transmit power control.

Another important parameter is the channel access probability $p_{tx}$ at each station, which, in the IEEE 802.11p system, can be controlled through information update frequency or message transmit rate $\lambda_F$, and the contention window size $CW_{min}$. The latter mainly limits the maximum value of $p_{tx}$. In general, the value of $p_{tx}$ shall stay small, as far as the resulting $\overline{T_{UI}(d_{RX})}$ at specific $d_{RX}$ value satisfies the requirement of service. For example, the value of $\lambda_F$ shall be below 60 [Hz] to meet $\overline{T_{UI}(8)} \leq 1\ s$ with $CW_{min} = 63$, $P_S = 512$ [B], and $r = 640$ [m], as shown in Figure 12(b).

Smaller frame duration $L$ is always preferable than bigger ones. This can be achieved by using more efficient encoding schemes for the message content or through link adaptation, i.e. using a higher modulation and encoding scheme, if the physical layer performance permits this.

As evaluated in this section, the performance of CAM broadcast in the vicinity of a vehicle is considered acceptable for the CAS as specified in the standard [2], especially when the above discussed control mechanisms are employed.

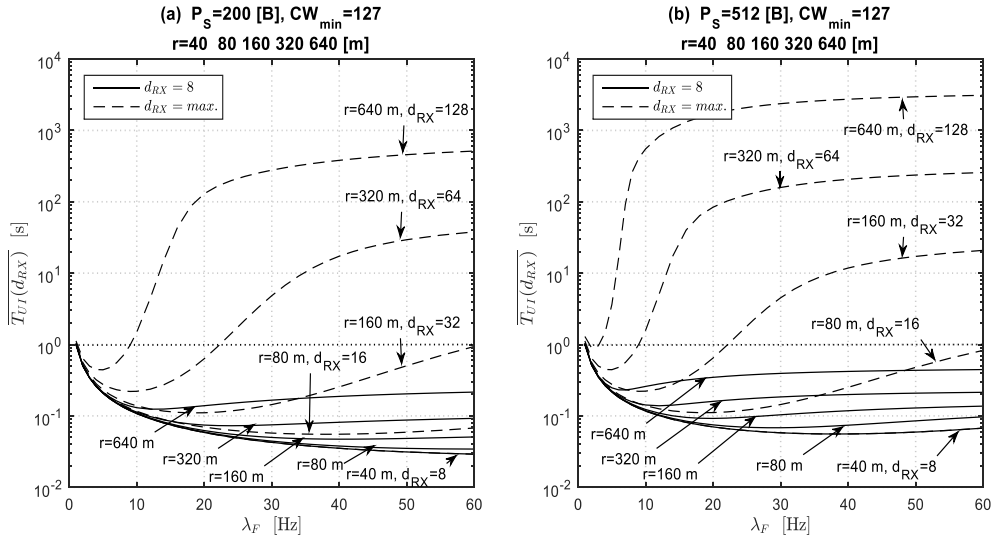

Figure 13: CAM update interval evaluated at $d_{RX}=8$ and $d_{RX}=$max. with $CW_{min}=127$

## VIII. CONCLUSIONS

A combination of the protocol model for IEEE 802.11p DCF originally developed in [3] and the hidden station model developed in [18] as introduced in this paper provides accurate analytical results for IEEE 802.11p radio broadcast with hidden stations under both, saturated and non-saturated traffic load. Besides reliability and efficiency performance, our analytical results also provide mean MAC layer access delay. The new models, quantitatively, provide how the contention window size of IEEE 802.11p and the traffic load at each station determine system performance for given parameters $L$ and $R$ in a hidden station scenario.

Analysis of the cooperative awareness service in highway scenarios based on our analytical models reveals how reliability of CAM broadcast deteriorates under increased vehicle density. For a given vehicle density, the reliability of CAM broadcast deteriorates with increased topological distance $d_{RX}$ between transmitter and receiver. However, depending on the value of $d_{RX}$, particularly for a close transmitter to receiver distance, e.g. $d_{RX} \leq 8$, the mean update interval below 1 [s] for CAM can be met. Our analytical model provides quantitative guidance on how to improve system performance in hidden station scenarios using control mechanisms like transmit power control, transmit rate control and link control.

## IX. REFERENCES

[1] Access layer part ETSI TS 102 687; Intelligent Transport Systems (ITS); Decentralized Congestion Control Mechanisms for Intelligent Transport Systems operating in the 5 GHz range. ETSI, 07 2011.

[2] ETSI EN 302 637-2, Intelligent Transport Systems (ITS); Vehicular Communications; Basic Set of Applications; Part 2: Specification of Cooperative Awareness Basic Service, 09 2014.

[3]  G. Bianchi. Performance analysis of the IEEE 802.11 distributed coordination function. *IEEE Journal on Selected Areas in Communications*, 18(3):535 – 547, Mar 2000.

[4]  Sheng-Tzong Cheng and Mingzoo Wu. Performance evaluation of ad-hoc WLAN by M/G/1 queueing model. In *Proc. Int. Conf. Information Technology: Coding and Computing ITCC 2005*, volume 2, pages 681–686, 2005.

[5]  P. E. Engelstad and O. N. Osterbo. Delay and Throughput Analysis of IEEE 802.11e EDCA with Starvation Prediction. In *The IEEE Conference on Local Computer Networks, 30th Anniversary*, pages 647–655, 2005.

[6]  Y. P. Fallah, Ching-Ling Huang, Raja Sengupta, and H. Krishnan. Analysis of Information Dissemination in Vehicular Ad-Hoc Networks with Application to Cooperative Vehicle Safety Systems. *IEEE Transactions on Vehicular Technology*, 60(1):233–247, 2011.

[7]  Michele Garetto, Jingpu Shi, and Edward W. Knightly. Modeling media access in embedded two-flow topologies of multi-hop wireless networks. In *Proceedings of the 11th Annual International Conference on Mobile Computing and Networking*, MobiCom '05, pages 200–214, New York, NY, USA, 2005. ACM.

[8]  R. Jennen. *Voice over IP Capacity of IEEE 802.11 WLAN and IEEE 802.21 based Interworking Performance of WLAN and Mobile LTE Networks*. PhD thesis, Communication Networks (ComNets) Research Group, RWTH Aachen University, 2012. URL: http://www.comnets.rwth-aachen.de/publications/dissertations.html.

[9]  L. Kleinrock. *Queueing Systems, Volume I: Theory*. Wiley Interscience, New York, 1975.

[10]  Xiaomin Ma. On the reliability and performance of real-time one-hop broadcast MANETs. *Wireless Networks*, 17:1323–1337, July 2011.

[11]  David Malone, Ken Duffy, and Doug Leith. Modeling the 802.11 Distributed Coordination Function in Nonsaturated Heterogeneous Conditions. *IEEE/ACM Transactions on Networking*, 15(1):159–172, 2007.

[12]  Sundar Subramanian, Marc Werner, Shihuan Liu, Jubin Jose, Radu Lupoaie, and Xinzhou Wu. Congestion control for vehicular safety: synchronous and asynchronous MAC algorithms. In *Proceedings of the ninth ACM international workshop on Vehicular inter-networking, systems, and applications*, VANET '12, pages 63–72, New York, NY, USA, 2012. ACM.

[13]  I. Tinnirello, G. Bianchi, and Yang Xiao. Refinements on IEEE 802.11 Distributed Coordination Function Modeling Approaches. *IEEE Transactions on Vehicular Technology*, 59(3):1055–1067, 2010.

[14]  M. Treiber, A. Kesting, and C. Thiemann. *Traffic Flow Dynamics: Data, Models and Simulation*. Springer Berlin Heidelberg, 2012.

[15]  Y. Xiao. Enhanced DCF of IEEE 802.11e to support QoS. In *IEEE Wireless Communications and Networking (WCNC 2003)*, volume 2, pages 1291 – 1296, Mar 2003.

[16]  Yuan Yao, Lei Rao, and Xue Liu. Performance and reliability analysis of IEEE 802.11p safety communication in a highway environment. 62(9):4198–4212, 2013.

[17]  Y. Zang. *Analysis of CSMA Based Broadcast Communication in Vehicular Networks with Hidden Stations*. PhD thesis, Aachen, Germany, Sep 2015. URL: http://www.comnets.rwth-aachen.de/publications/dissertations.html.

[18]  Yunpeng Zang, Bernhard Walke, Guido Hiertz, and Christian Wietfeld. CSMA-based Packet Broadcast in Radio Channels with Hidden Stations. *Submitted to IEEE Transactions on Wireless Communications*, 2016. URL: https://arxiv.org/abs/1612.03448 .